# Monolithic Integration of Embedded III-V Lasers on SOI

Wen-Qi Wei[1,3†], An He[2†], Bo Yang[1†], Jing-Zhi Huang[1,4], Dong Han[1], Min Ming[1], Zi-Hao Wang[1,3,4], Xuhan Guo[2,*], Yikai Su[2,*], Jian-Jun Zhang[1,3,4,*], Ting Wang[1,3,4,*]

[1]Institute of Physics, Chinese Academy of Sciences, Beijing, China
[2]State Key Laboratory of Advanced Optical, Communication Systems and Networks, Department of Electronic Engineering, Shanghai Jiao Tong University, Shanghai, China
[3]Songshan Lake Materials Laboratory, Dongguan, Guangdong, China
[4]School of Physical Sciences, University of Chinese Academy of Sciences, Beijing, China
*Corresponding authors: guoxuhan@sjtu.edu.cn, yikaisu@sjtu.edu.cn, jjzhang@iphy.ac.cn, wangting@iphy.ac.cn
[†]These authors contribute equally to this work.

Silicon photonic integration has gained great success in many application fields owing to the excellent optical device properties and complementary metal-oxide semiconductor (CMOS) compatibility. Realizing monolithic integration of III-V lasers and silicon photonic components on single silicon wafer is recognized as a long-standing obstacle for ultra-dense photonic integration, which can provide considerable economical, energy efficient and foundry-scalable on-chip light sources, that has not been reported yet. Here, we demonstrate embedded InAs/GaAs quantum dot (QD) lasers directly grown on trenched silicon-on-insulator (SOI) substrate, enabling monolithic integration with butt-coupled silicon waveguides. By utilizing the patterned grating structures inside pre-defined SOI trenches and unique epitaxial method via molecular beam epitaxy (MBE), high-performance embedded InAs QD lasers with out-coupled silicon waveguide are achieved on such template. By resolving the epitaxy and fabrication challenges in such monolithic integrated architecture, embedded III-V lasers on SOI with continuous-wave lasing up to 85 °C are obtained. The maximum output power of 6.8 mW can be measured from the end tip of the butt-coupled silicon waveguides, with estimated coupling efficiency of approximately -7.35 dB. The results presented here provide a scalable and low-cost epitaxial method for realization of on-chip light sources directly coupling to the silicon photonic components for future high-density photonic integration.

There is fast-growing demand for integrated silicon photonic chips incorporating with on-chip lasers, which can lead to power-efficient and densely integrated optical interconnects and high-speed optical communications[1-4]. Various external lasers coupled silicon integrated chips have been demonstrated on 200 mm silicon-on-insulator (SOI) wafers, however, on-chip lasers remain as the stumbling block for further development of ultra-dense silicon photonic integration[5,6]. Over the past decade, III-V/Si heterogeneous integration via bonding techniques is academically and commercially recognized as a promising path towards realization of on-chip light sources[7-10]. With rapid progresses of silicon integrated photonics in applications such as artificial intelligence, hyper-scale data centers, high-performance computing, light-detection and ranging (LIDAR) and microwave photonics[11], monolithically integrated light sources begin to show their increasing preference as an alternative technical trend for compactness and low-power consumption[12].

As monolithic integration of III-V laser and silicon photonic components has always been a much-desired functionality, direct epitaxial growth of III-V quantum-dot (QD) lasers on silicon substrate has been extensively investigated with dramatic progress recent years[13-20]. Benefiting from the technical development of high-quality III-V material growth on silicon[21-25], various silicon-based laser structures have been demonstrated with outstanding performance, including distributed feedback (DFB) lasers[26-29], microcavity lasers[30-32] and mode-locked lasers [33-36]. From silicon photonic integration perspectives, all active and passive silicon photonic components should be on the same SOI platform, therefore, there is a strong urge to develop an SOI-based monolithic laser integration solution to efficiently couple the light into silicon waveguides (WGs). Up to date, there are relatively limited researches focusing on direct growth of SOI-based lasers [37-40], and usually the lasers are located on top silicon of SOI with thick III-V buffer layers which are not capable of being coupled with passive waveguides. In case of heterogeneous bonding techniques mentioned above, the coupling from bonded lasers to silicon WGs has been well established by evanescent coupling[7]. On the other hand, although directly grown III-V lasers have been systematically investigated, the monolithic integration between III-V lasers and silicon photonic components remains absent in the field. In this work, we have demonstrated the first embedded QD lasers on SOI substrate with pre-patterned laser trenches and silicon waveguides, with monolithic coupling to silicon passive waveguides, which offers great potential of achieving fully monolithic silicon photonic integrated chip.

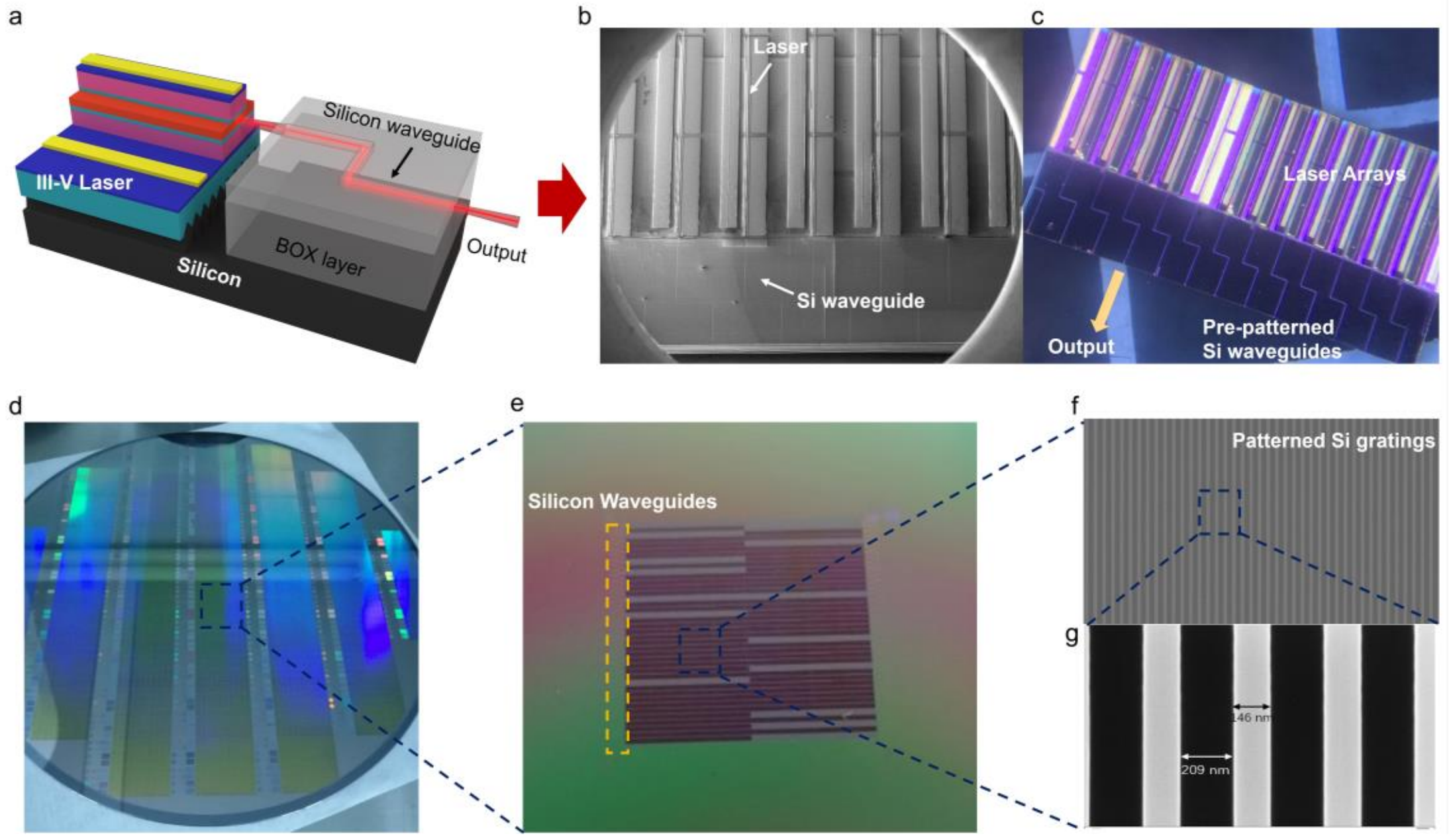

**Fig. 1 Monolithically integrated embedded InAs QD lasers on SOI characterizations in this work. a** Schematic of monolithic integration of III-V QD laser edge coupled with silicon waveguide on SOI platform. **b** Top-view SEM image of InAs QD laser arrays grown in pre-patterned laser trenches, with passive silicon WGs. **c** Optical microscope image of entire integrated chip. **d** 8-inch SOI wafer with pre-patterned laser trenches and silicon waveguides. **e** Zoomed-in optical microscope image of laser trenches with aligned silicon waveguide arrays on SOI substrate. **f** Top-view SEM image of patterned silicon grating structures inside laser trenches for III-V growth. **g** Magnified silicon gratings with slab width of 146 nm and gap width of 209 nm. The duty cycle of this grating is approximately 40% inside the laser trench.

Fig. 1a depicts the schematic of monolithically integrated III-V QD lasers edge-coupled with silicon waveguides on SOI substrate. The scanning electron microscope (SEM) image and optical microscope image of fabricated devices are shown in Fig. 1b and c. Starting with an 8-inch SOI wafer, silicon waveguides are pre-patterned on top silicon layer of SOI substrate (Fig. 1d). The laser trench is then produced via dry etching process through buried oxide (BOX) layer into bulk silicon substrate for III-V laser growth, as shown in Fig. 1e. Inside laser trench, periodic silicon grating structures are then patterned with duty cycle of approximately 40% (146 nm slab width with 209 nm gap) as the surface SEM images show in Fig. 1f. The grating structures cover the entire laser trench for high quality III-V direct epitaxial growth. The zoomed-in SEM image of silicon gratings is displayed in Fig. 1g.

## Results and discussion

**Design and fabrication of SOI template.** Fig. 2a presents the fabrication process of the SOI devices. The edge coupler and patterned trench are manufactured on an SOI wafer with a 220 nm thick top Si layer and a 3 μm thick buried $SiO_2$ layer. The fork-shaped coupler and interconnecting waveguide are defined through E-beam lithography (EBL) process. The resist pattern was fully etched using the ICP-RIE process. Subsequently, 3.5 μm thick $SiO_2$ cladding layer is deposited by plasma-enhanced chemical vapor deposition (PECVD) after removing the e-beam photoresist. The detailed fabrication information of the edge coupler can be found in the Methods. Then, the cladding, top Si, BOX, and 1.5 μm substrate layer are etched to form the laser trench. Finally, the silicon gratings are fabricated by EBL and ICP-RIE etching for III-V laser growth.

Compared to common inverse taper edge coupler with single tip, edge couplers with multiple taper tips could have higher coupling efficiency and better alignment tolerance for laser butt-coupling scheme due to its elliptical spotsize are more comparable to laser's mode profile[41,42]. Hence, considering both eminent performance and tractable fabrication process, fork-shape edge couplers with double tips are proposed[43,44]. Besides the tip width, the fork shape offers an extra design parameter, while the gap between two tips can be also used to expand the mode field.

The structure of the fork-shape coupler is shown in Fig. 2b. The same design methods are followed from our previous work[41,42] with further optimized structures. In order to verify the performance of the edge couplers used in monolithically integrated embedded InAs QD lasers on SOI wafer, we separately fabricate the silicon edge couplers with the same parameters and measured with a single mode fiber (SMF) and InAs QD laser grown on silicon substrate via butt coupling desgin. The tip width (w) is set as 100 nm, and the gap (g) between the two tips is 3.4 μm. The length of the first fork stage (L1) which converts the fiber mode to the slot waveguide mode is 64 μm. The second fork stage, converting the slot waveguide mode to the strip waveguide mode, is 11 μm long. Here, Fig. 2c shows the simulation result of electric field distribution of this edge coupler. The SMF or the monolithically integrated laser modes, which both exhibit relatively large mode profile, transfer preferentially into the edge coupler facet with large modal overlap, and then adiabatically converts to the strip waveguide mode. The edge coupling loss is mainly determined by the mode similarity, which is also called mode overlap, between the light input and the coupler facet[41]. The mode overlap between the SMF with mode field diameter (MFD) of 10 μm and the chip facet is 85%. In comparison, the mode overlap between InAs QD laser and fork shape coupler is approximately 76%, which is relatively lower than SMF. Notably, the total length of the coupler is only 75 μm, which is superior for precious integrated photonic chip.

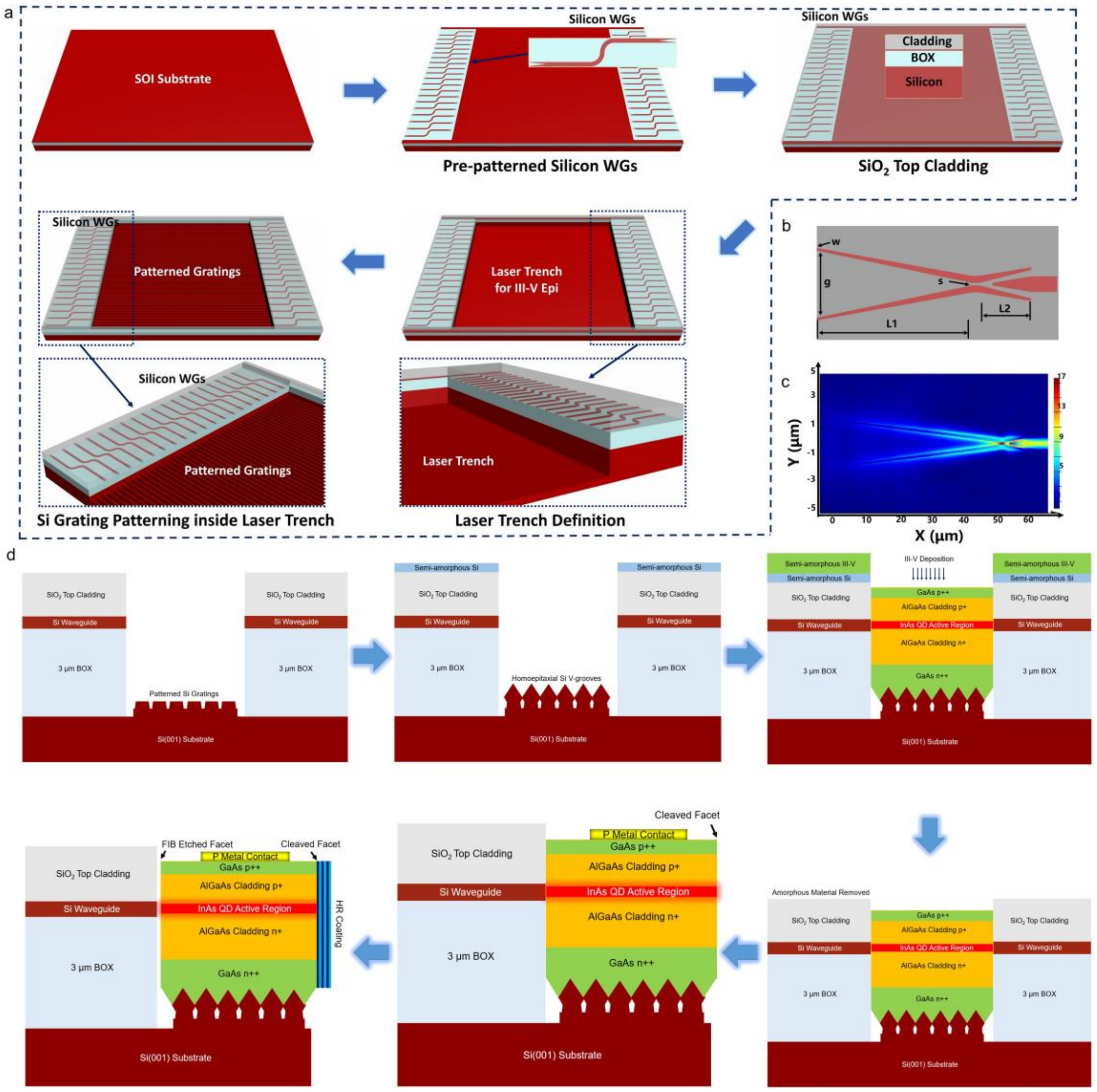

**Fig. 2 Schematic diagram of patterned SOI trenches and design of edge coupler. a** Fabrication flows of patterned SOI template with 3 μm BOX layer and 220 nm top Si layer. **b** The layout and parameters of silicon fork coupler. **c** Electric field distribution of the edge coupler. **d** Schematic diagram of embedded laser process on trenched SOI. Step 1: The exposed silicon substrate patterned with silicon gratings. Step 2: Homoepitaxially formed Si V-groove structures over the top of silicon gratings. Step 3: InAs/GaAs QD laser epi-structures directly grown inside the SOI trench. Step 4: Chemical remove of unwanted III-V materials outside the SOI trench. Step 5: Fabricated narrow ridge laser with one-side as-cleaved facets. Step 6: Final embedded QD lasers with direct edge coupling into pre-patterned silicon waveguides.

**Monolithic epitaxial growth and fabrication of III-V lasers on trenched SOI substrate.** Fig. 2d shows the entire monolithic process of on-chip lasers, including the embedded growth of III-V gain materials in the grating patterned SOI trench and subsequent laser fabrications. As depicted in Fig. 2d, the pre-patterned silicon waveguides are protected with PECVD $SiO_2$ top cladding. The SOI template with pre-defined silicon waveguides is firstly prepared in order to create the laser trench on bottom silicon of SOI substrate for laser material deposition. To note, as both the horizontal and height alignments to silicon waveguides are critical, the depth of laser trench needs to be precisely designed in line with the active region height of InAs QD laser on silicon. Although, the exact alignment can still be finely adjusted during the epitaxial growth, which normally can be controlled within accuracy of nanometer scale. In order to avoid anti-phase domains (APDs), threading dislocations (TDDs) and thermal mismatch induced thermal cracks[45,46], homoepitaxially formed (111)-faceted Si V-grooves over the top of pre-defined silicon gratings are introduced here to suppress defects generated during hetero-epitaxial growth as shown in step 1 and 2 of Fig. 2d. These techniques will been discussed explicitly in the Methods, while the entire epitaxial structures will be described in the following section.

There is one major issue here as observed in step 3 of Fig. 2d, during the hybrid III-V/IV growth process via molecular beam epitaxy, both silicon and III-V materials will also be deposited outside the laser trench in semi-amorphous crystal format. In this case, it normally leads to relatively large height contrast in- and outside of the laser trench, which can result in uneven photoresist stacking at the edge of the trench during laser-waveguide alignment process. To solve this problem, $H_3PO_4$: $H_2O_2$: $H_2O$ (1 : 2 : 20) wet etching solution is selected here to remove excessive semi-amorphous III-V materials outside the trench region before the laser epitaxy process.

The embedded III-V laser is then processed with one-side cleaved facet (step 5 of Fig. 2d). As previously mentioned, the alignment precision between laser and silicon waveguide will significantly affect coupling efficiency especially in the vertical direction. Meanwhile, the accurate control of alignment in horizontal direction is determined by photolithograph during the laser device process. The alignment deviation between the central axis of silicon waveguide and the laser ridge here can be controlled within ±250 nm, which is within the tolerance of the unique multipe tips taper design. To ensure precise alignment between the laser ridge and silicon waveguide, the removal process of semi-amorphous silicon and III-V materials above the waveguide has to be treated properly, otherwise, the blurred silicon waveguide structure will lead to reduced accuracy of laser-to-waveguide alignment. In addition, the semi-amorphous material removing process mentioned above acts also as facet etching for III-V/Si interface at laser output side, as the gap area between laser and silicon waveguide remains un-protected during the removing process.

The integrated device is finalized by applying high reflection coating to as-cleaved facet at one side, while implementing focused ion beam (FIB) milling process to the other side for the laser output facet as shown in step 6 of Fig. 2d. The ion milling process here aims to create mirror-like facet by removing the semi-amorphous material at III-V/IV interface. Although the embedded laser can still lase without FIB treatment, the facet polishing using FIB can further improve the laser performance significantly. The performance differences between wet etched facets and FIB etched facets will be compared and discussed in the laser characterization section.

In case of the monolithic epitaxy of embedded III-V lasers, the growth details are described in Methods. The overall schematic of the laser epi-structure on the trenched SOI template is displayed in Fig. 3a. Here, a 10 nm-thick AlAs nucleation layer was first deposited to optimize the GaAs/Si (111) interface and suppress APD formation as marked in Fig. 3a. After approximately 2100 nm-thick III-V buffer layers, which includes InGa(Al)As/GaAs quantum well dislocation filters (DLFs) and GaAs/AlGaAs superlattice (SL) layers[24], a smooth and APD-free GaAs surface can be achieved on the trenched SOI region as verified by atomic force microscope (AFM) measurement shown in Fig. 3b. The $5 \times 5\ \mu m^2$ AFM image shows the root-mean-square (RMS) roughness of only 0.8 nm. Fig. 3c shows the surface electron channeling contrast imaging (ECCI) result of the trenched GaAs/SOI template, indicating a $2.6 \times 10^7/cm^2$ surface threading dislocation density (TDD) of the template. The bright-filed cross-sectional transmission electron microscopy (TEM) image at the GaAs/Si(111) interface is also presented in Fig. 3d, which indicates the same defect suppression effect of the homo-epitaxially formed Si(111) sawtooth structures reported previously[32,34].

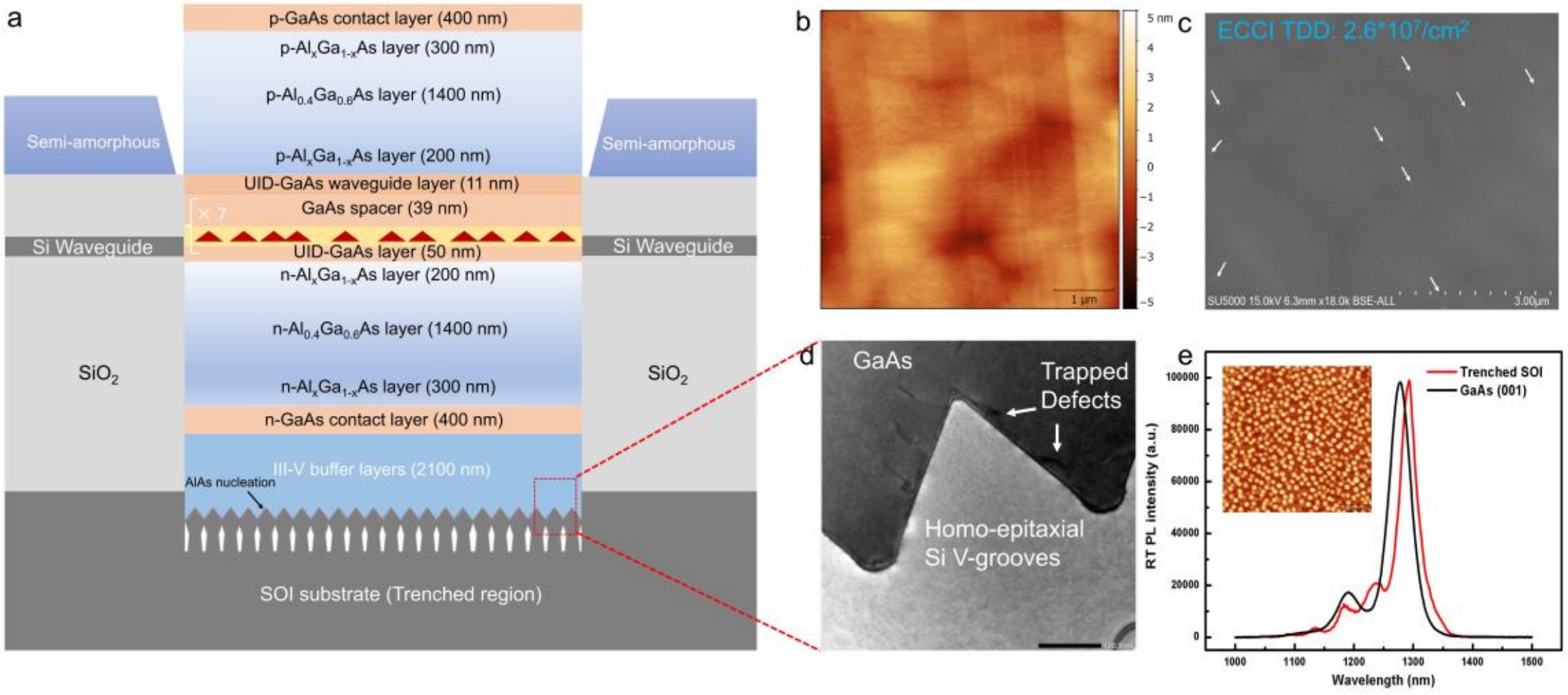

**Fig. 3 Epitaxial growth structures and material characterization. a** Schematic of the laser epi-structures. **b** Surface AFM image of 2100 nm thick III-V buffer layers grown on trenched SOI before epitaxial growth of laser sturctures (RMS ~ 0.8 nm). **c** TDD estimated from ECCI image (2.6 X $10^7$ /$cm^2$). **d** Cross-sectional TEM image of as-grown GaAs/Si interface on trenched region of SOI substrate. **e** PL spectra comparison between InAs QDs grown on trenched SOI substrate and standard GaAs substrate under identical conditions. Inset: AFM image of surface InAs QDs on trenched SOI.

Based on these high-quality trenched GaAs/SOI templates, the standard InAs/GaAs QD laser diode structures[39,47] are grown as shown in Fig. 3a. The laser structure consists of a 7-layer InAs QD active region, which is sandwiched between the 400 nm n-/p-doped GaAs contact layers and 1400 nm n-/p-doped $Al_{0.4}Ga_{0.6}As$ cladding layers. Astride each AlGaAs cladding layer, the step-graded $Al_xGa_{1-x}As$ ($0.1<x<0.4$) transitional layers are deposited in order to increase the current injection efficiency of the device. As Fig. 3a shows, the semi-amorphous III-V materials will be formed on the $SiO_2$ cladding layers outside the trenched region, which can be removed by wet-etching process before device fabrication as mentioned above. To clarify the optical gain properties of the InAs QDs on the trenched GaAs/SOI substrate, identical 7 layers of InAs QDs are both grown on the trenched GaAs/SOI and GaAs (001) substrates, respectively, and the room-temperature photoluminescence (PL) spectra of the two samples are shown in Fig. 3e. Typical O-band PL emission with a full width at half maximum (FWHM) of 33 nm is obtained from the InAs QDs on trenched GaAs/SOI, which is interestingly smaller than that (FWHM: 42 nm) on GaAs (001) substrate. Almost the same PL peak intensity from the two samples is observed here. The inset in Fig. 3e shows the $1 \times 1\ \mu m^2$ AFM image of the surface InAs QDs on trenched GaAs/SOI, indicating a $5.1 \times 10^{10}/cm^2$ dot density and good dot uniformity. Notably, the offset in the PL peaks from the two samples (trenched SOI: 1293 nm; GaAs: 1278 nm) is caused by the difference of the real temperatures on the two substrates, which will be discussed in Methods.

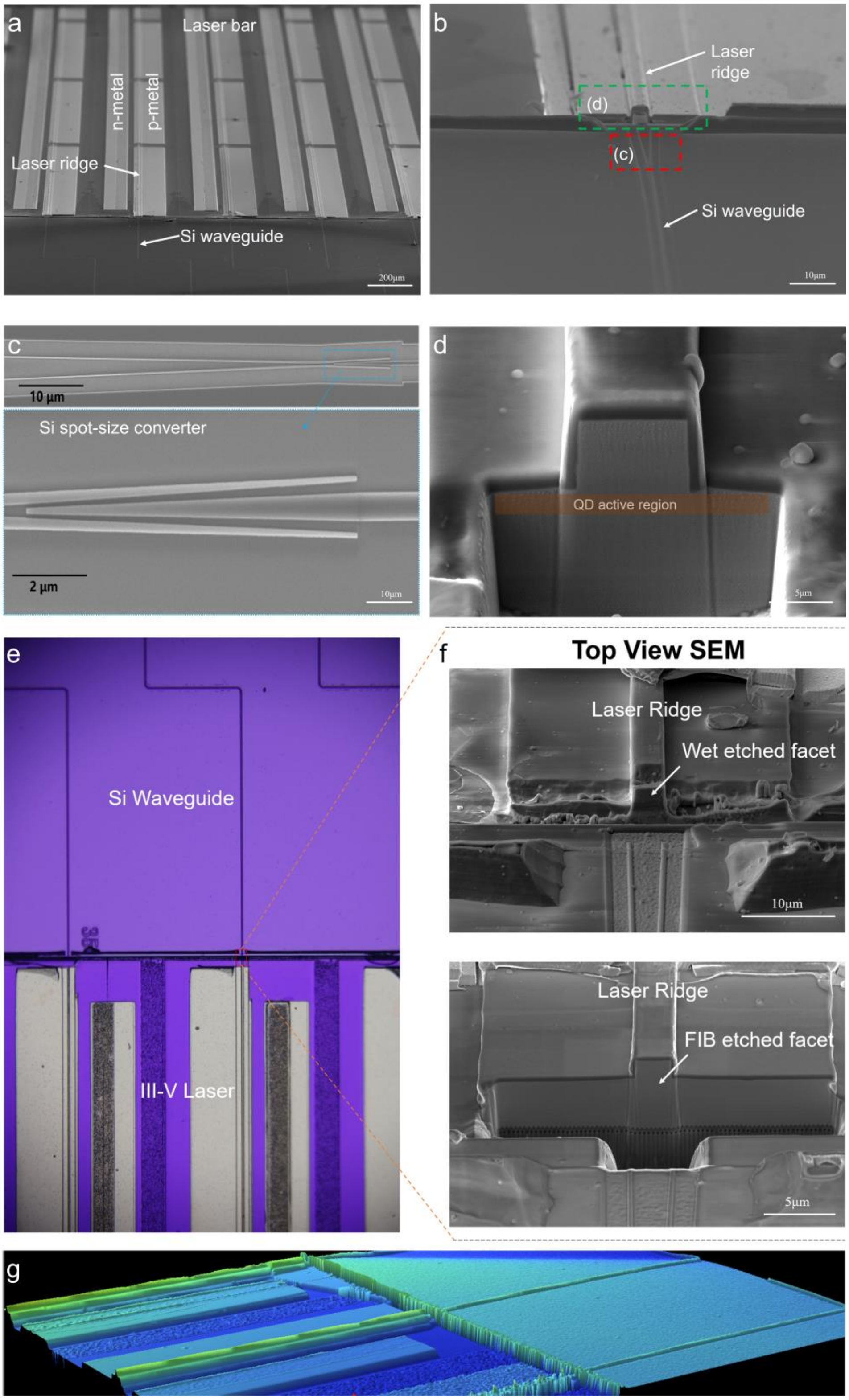

**Fig. 4 Fabricated monolithically integrated InAs QD lasers coupled to silicon waveguides. a** Tilted SEM image of entire integrated chip. **b** SEM image of fabricated narrow ridge laser directly coupled with silicon waveguides using fork-like mode converter. **c** SEM image of silicon waveguide mode converter. **d** Magnified SEM image of single embedded laser edge-coupled with silicon waveguide. **e** Optical microscope of monolithically integrated laser and silicon WG. **f** SEM images of the embedded lasers with wet etched and FIB etched facets, respectively. **g** White light interferometric image of the finalized

devices.

Fig. 4a shows the tilted-view SEM image of fabricated embedded InAs QD lasers with precise alignment to pre-patterned silicon WGs. As previously mentioned, the horizontal offset between laser ridges and silicon waveguides is less than 500 nm along with the arrays of waveguide lasers (Fig. 4b). At the coupling tip of silicon WGs, fork-like spot-size converter is implemented to increase dimensional tolerance of laser-waveguide alignment offsets, as shown in Fig. 4c. As the laser facet close to silicon waveguide intends to accumulate some semi-amorphous materials, the wet etched facet is normally difficult to achieve mirror-like sidewall. FIB milling is also utilized here to further polish the facet for high gain cavity formation as shown in Fig. 4d. Fig. 4e shows the zoomed-in optical microscope image of the monolithically integrated device. The smoothness of the cavity facet is one of the most essential factors that affect the laser performance. Therefore, the FIB etch with large current is implemented to initially separate the III-V and silicon materials. Smaller current FIB is then applied to the sidewall for fining polishing, in order to obtain shinning laser facet. Here, we compare the wet etched laser facet with FIB polished laser facet in Fig. 4f, where the upper SEM image shows the wet etched laser facet with approximately 5 μm wide coupling gap between laser and silicon WGs. As observed, both steepness and roughness of the wet etched laser facet are imperfect. After FIB fine polishing in the lower SEM image of Fig. 4f, the laser facet appears to be ultra-smooth that is similar to as-cleaved facet. In Fig. 4g, white light interferometric imaging is performed to show the entire structure of integrated device, where the left side is the embedded InAs QD laser, and the right side is the pre-patterned silicon waveguides.

**Characterizations of on-chip integrated lasers.** In order to examine the effective coupling efficiency from laser to silicon WGs, here, we select single laser bar directly grown inside SOI trench with both facets cleaved as a reference laser, which is characterized with standard light-current (L-I) measurements (top left of Fig. 5a). Here, the reference laser is produced as an analogy to InAs QD laser directly grown on silicon substrate. In case of silicon WG coupled on-chip laser, both L-I curves and optical spectra are collected from the silicon WG side as shown in Fig. 5a. The temperature dependent L-I curves of embedded laser without silicon WG (so-named as reference laser) are measured in Fig. 5b, which manages to lase up to 95 $^{o}$C in continuous-wave (CW) current operation. The threshold current at room temperature is approximately 50 mA. The maximum output power is 37 mW at injection current of 250 mA.

The silicon WG edge-coupled on-chip laser is then characterized with slightly higher threshold current of 65 mA at room temperature under CW mode, while the maximum operation temperature is reduced by 10 $^{o}$C to 85 $^{o}$C (Fig. 5c). The relatively higher threshold current and lower maximum operation temperature are attributed to increased thermal accumulation inside the laser trench with surrounding BOX layer. The L-I performance of embedded laser with single-side wet etched facet and FIB etched facet are compared in inset of Fig. 5c. With additional FIB facet polishing, the threshold current is reduced from 92 mA to 65 mA, while the out-coupled optical power through silicon WG is also improved from 5.3 mW to 6.8 mW at injection current of 210 mA. Due to relatively large divergence angle of InAs QD laser and slightly height mismatch induced during the gain material growth, the detected output power from silicon WG side is lower than the actual laser power. The overall coupling loss is here estimated to be -7.35 dB. Furthermore, in order to extract characteristic temperature ($T_o$) and slope efficiency of both as-cleaved laser and on-chip integrated laser, threshold current and output power at different temperatures are analyzed as shown in Fig. 5d. The as-cleaved laser on trenched SOI exhibits $T_o$ values of 282.4 K and 51 K in temperature ranges from 20 - 40 °C and 45 - 95 °C. In comparison, on-chip integrated laser (silicon WG output) has slight degradation in $T_o$ values, that are 108.1 K and 33.7 K in temperature ranges from 20 - 65 °C and 70 - 85 °C. Moerover, as Fig. 5c shows, the reference laser presents a higher slope efficiency of 0.148 W/A at 20 °C, while the silicon WG coupled on-chip laser owns that of 0.025 W/A at 20 °C, which are induced from non-optimized edge-coupling efficiency and slightly height mismatch induced misalignment between embedded laser and silicon WG. Fiber collimator is here implemented to collect light from silicon WG side for optical spectral analysis. Current dependent spectral measurements from 70 mA to 160 mA at room temperature are shown in Fig. 5e. The spectral evolution of on-chip integrated laser operating in the temperature ranging from 20 °C to 70 °C at fixed injection current of 175 mA is measured in Fig. 5f. To the best of our knowledge, this is the first demonstration of the InAs/GaAs QD laser was epitaxially grown on a trenched SOI template with a butt-coupling silicon waveguide.

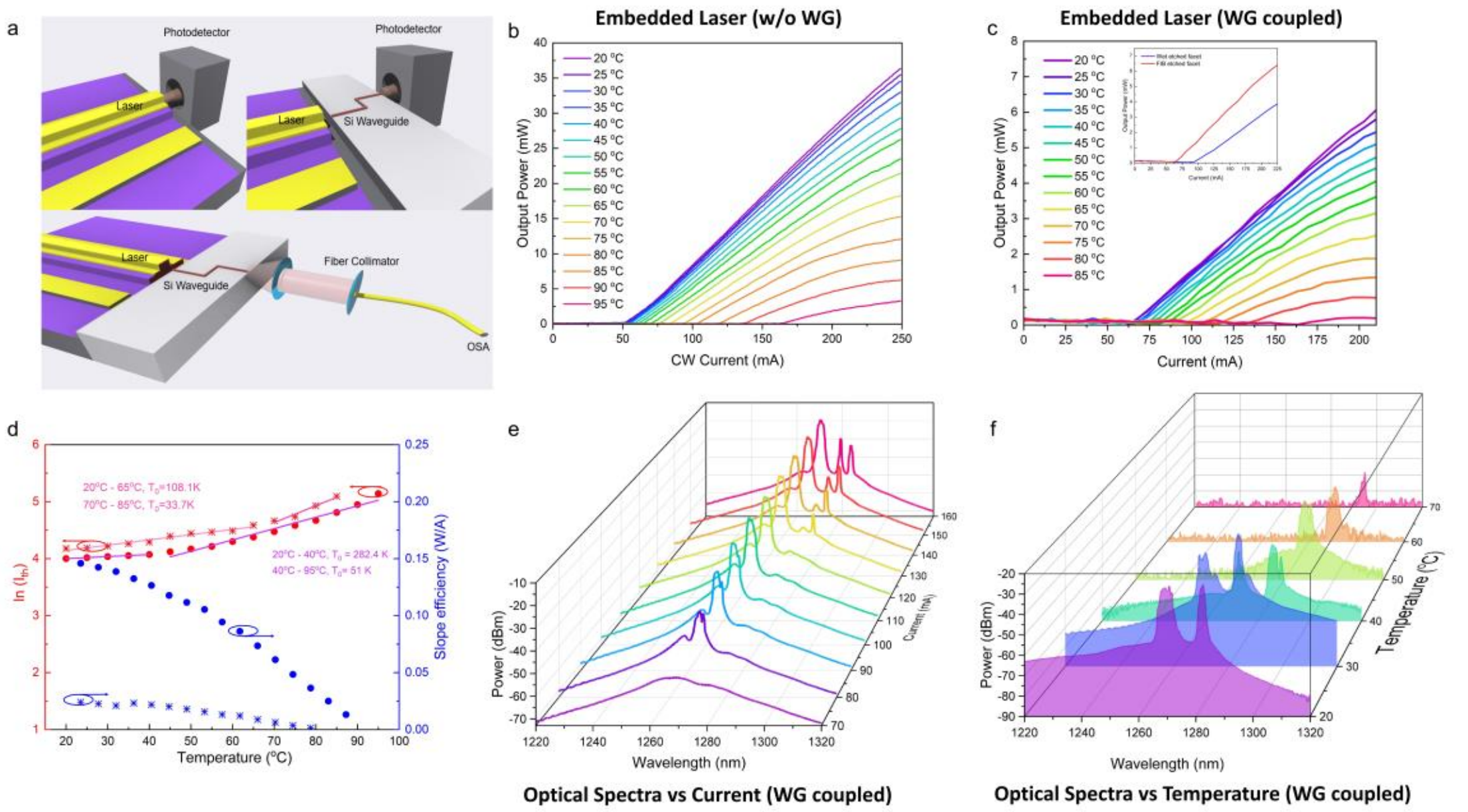

**Fig. 5 Continuous-wave characterizations of embedded InAs QD laser on SOI with and without coupling into silicon waveguide. a** Schematics of L-I and optical spectral measurements; top left: L-I measurements of double-side cleaved III-V laser inside SOI trench; top right: L-I measurements of embedded InAs QD laser from silicon waveguide output; bottom: optical spectral measurements from the silicon waveguide. **b** Continuous-wave temperature-dependent L-I measurements of double-side cleaved III-V laser inside SOI trench as reference laser. **c** Continuous-wave temperature-dependent L-I measurements of integrated laser with one cleaved facet and FIB etched for the other; inset: room-temperature L-I comparison between single-side wet etched facet and FIB etched facet. **d** Plots of natural logarithm of threshold current and slope efficiency versus varied operating temperatures; characteristic temperature ($T_o$) are fitted for double-side cleaved embedded laser (red dots) in temperature ranges from 20 - 40 °C and 45 - 95 °C, respectively; characteristic temperature ($T_o$) are fitted for embedded laser with integrated silicon WG (red stars) in temperature ranges from 20 - 65 °C and 70 - 85 °C, respectively. **e** Optical spectral analysis of integrated laser versus increased injection current. **f** Optical spectral analysis of integrated laser versus temperature variation.

In summary, monolithic integrated III-V lasers on SOI substrate with silicon waveguide output have been realized by directly growing InAs QD lasers inside pre-patterned SOI trenches. Homoepitaxial formation of (111)-faceted Si V-grooves and heteroepitaxial growth of InGaAs/GaAs defect trapping techniques are implemented in this work to achieve high quality III-V gain materials on SOI. Our results demonstrate that monolithic integration of III-V laser with silicon photonic components will no longer be a design-level hypothesis. Overall, the monolithically integrated lasers can operate over 85 °C with low threshold current of 65 mA at room temperature and silicon WG coupled maximum output power of 6.8 mW. One more step forward, the performance of on-chip integrated InAs QD lasers can be further improved by including advanced silicon spot size converter with accurate control of laser-waveguide coupling distance during process. Once the coupling efficiency issue is resolved, many selections of silicon photonic components can all be integrated monolithically on a single wafer, such as modulators, wavelength de-multiplexers and photodetectors, just to name a few. We believe that this monolithic integration techniques of on-chip lasers would offer a promising approach towards high-density and large-scale silicon photonic integration, especially in the application fields such as on-chip optical interconnect and integrated optical ranging.

# Methods

**Fabrication of silicon edge couplers.** The fork shape coupler and interconnecting waveguide are defined through a single E-beam lithography (EBL) process with Vistec EBPG-5200+ Electron-beam lithography system. The patterns on 400 nm thick AR-P 6200.09 photoresist are fully etched using the SPTS ICP Deep Silicon Etching System, the flow of $C_4F_8$ and SF6 are 45 sccm and 20 sccm, respectively. Subsequently, after removing the e-beam photoresist, a 3 μm thick $SiO_2$ cladding layer is deposited by plasma-enhanced chemical vapor deposition (PECVD) process with OXFORD Plasmalab System 100, at 300 °C temperature. The cladding, top Si and BOX layers are etching with NMC ICP Reactive Ion Etching System, the flow of $CHF_3$ and Ar are 80 sccm and 35 sccm, respectively. Afterwards, 1.5 μm depth of the substrate Si layer are etched with SPTS ICP Deep Silicon Etching System to form a trench. Finally, the gratings are fabricated on the substrate in the trench by EBL and ICP etching for the laser growth.

**Material growth.** The epitaxial materials were prepared by our dual chamber solid-source MBE system (hybrid III-V/IV) to allow *in-situ* transfer between III-V and IV chamber without exposing to atmosphere. The SOI substrates are first diced into 3.2×3.2 $cm^2$ pieces, which are suitable for the 4-inch MBE system, and cleaned with standard RCA process to remove the surface contaminants. Before loading into MBE chamber, the SOI template was dipped into diluted HF solution (HF : $H_2O$ = 1 : 10) strictly within 10 seconds to remove the natural oxide layer in the "U"-shape grating-patterned trenched region, while avoiding horizontal wet-etching of the BOX layers surrounding the trenched region and surface etching of $SiO_2$ top cladding layer. The epitaxial growth is initiated with 420 nm thick homoepitaxial growth of silicon buffer layer at substrate temperature of 600 °C with a 1 Å/s growth rate in Group IV MBE chamber, to construct uniform sawtooth structures with Si (111) Miller facets in the trenched SOI region. The substrate was then *in-situ* transferred into III-V growth chamber for the following growth of III-V buffer layers and InAs/GaAs QD laser structure. To note, although the geometry of the trenched SOI substrates with passive silicon WGs is much different from the initial SOI substrate, the disturbance in substrate growth temperature is still negligible in III-V epitaxial growth process. After *in-situ* transferred into GaAs MBE chamber, the substrate was heated to 360 °C for growth of 10 nm AlAs and 30 nm GaAs nucleation layers. Then, 560 nm GaAs buffer layer was grown at 540 °C with a 1 Å/s growth rate to flatten the unevenness caused by the sawtooth structures. One repeat of 10 times 10 nm-$In_{0.13}Ga_{0.87}As$/10 nm-GaAs quantum wells and two repeats of 10 times 10 nm-$In_{0.15}Al_{0.85}As$/10 nm-GaAs quantum wells were grown at 440 °C as dislocation filters, which are separated by 200 nm GaAs strain-relaxed buffer layers. After depositing 3 periods of $Al_{0.6}Ga_{0.4}As$/GaAs SLs, the trenched GaAs/SOI template with smooth surface and low TDD in trenched region was achieved.

On the trenched GaAs/SOI template, 400 nm thick heavily n-doped GaAs contact layer was first grown at 540 °C, following by 300 nm step-graded $Al_xGa_{1-x}As$ transitional layers, 1400 nm $Al_{0.4}Ga_{0.6}As$ cladding layer, and 200 nm step-graded upper $Al_xGa_{1-x}As$ transitional layers, which were all n-doped. The active region includes 7 layers of InAs QDs, which was sandwiched by 50 nm and 11 nm unintentionally doped (UID) GaAs layers as shown in Fig. 3a. Each of the InAs QD layers consists of 8.1 Å InAs QD, 4 nm $In_{14.8}Ga_{85.2}As$ capping layer and 39 nm GaAs spacer layer. The InAs QD and capping layers are grown at an optimized temperature of 420 °C, which is 20 °C less than that used on GaAs/Si (001) and GaAs (001) substrates. Above the QD active region, symmetric upper AlGaAs structures were grown at 560 °C but with p-type doping. At last, a 400 nm heavily p-doped GaAs contact layer was deposited at 540 °C with a 1 Å/s growth rate.

**Fabrication of on-chip integrated lasers.** The integrated lasers were fabricated from embedded growth of III-V gain materials on SOI described above by using the standard lithography and dry etching techniques. For the definition of 3 μm wide ridges, DWL66+ Laser Direct Writing system with SPR 220 photoresist is used here to allow flexibility of laser-waveguide alignment. The ridges are dry etched with Plasmalab System 100 ICP180 system ($BCl_3$: $Cl_2$: $Ar_2$ = 10: 6: 4) for 2.2 μm etching depth. P-type and N-type metal contacts are deposited as Ti/Pt/Ti/Au and Ni/Ge/Au/Ti/Au by using PEVA-600E electron beam evaporation system and DZ-300 thermal evaporation sedimentary system, respectively. Additional 2.3 μm thick inter-metal contact were deposited under three separate processes with different deposition angles (-45°, 45°, 0°), which exhibit thickness of 20/700 nm, 20/700 nm, and 20/900 nm, correspondingly.

After the completion of laser process, the substrate was thinned to 70 μm for single facet cleaving. The length of the laser cavity is 3 mm. The laser's facet that is away from the waveguide was cleaved to form laser cavity. HR coating was also applied to this facet. The HR coating consists of 8 pairs of $SiO_2$ (145 nm) and $Nb_2O_5$ (228nm) superlattice. The reflection of the HR coating is measured as 95 %.

## Data availability

The data underlying the results presented in this paper are not publicly available at this time but may be obtained from the authors upon reasonable request.

## Acknowledgements

We gratefully acknowledge support from National Key Research and Development Program of China (2021YFB2800400), National Natural Science Foundation of China (Grant No. 61975230, 62008308). Ting Wang was supported by the Youth Innovation Promotion Association of CAS (No. 2018011). Mr. Mingchen Guo, Mr. Xianbiao Hu and Mr. Zhen Li are acknowledged for their early contributions in device fabrications and material growth.

## Author contributions

Y. Bo fabricated the integrated laser device. A. He performed the design and fabrication of SOI template. W. Q. Wei and D. Han performed the direct epitaxial growth of III-V laser inside SOI trenches. Y. Bo, J. Z. Huang and Z. H. Wang performed the characterization of integrated lasers. M. Ming performed the focused ion beam facet polishing. T. Wang, H. Guo, Y. Su and J. J. Zhang provided overall technical leadership across all aspects of research.